\documentclass[floatfix, preprintnumbers, twocolumn, superscriptaddress, amsmath, amssymb, aps, prl]{revtex4-1}
\usepackage{graphicx, xcolor}
\usepackage[caption=false]{subfig}
\newcommand*\xbar[1]{%
   \hbox{%
     \vbox{%
       \hrule height 0.5pt 
       \kern0.4ex
       \hbox{%
         \kern-0.1em
         \ensuremath{#1}%
         \kern-0.2em
       }%
     }%
   }%
}%
\usepackage[utf8]{inputenc}
\usepackage{soul}
\usepackage{times}
\usepackage{multirow}
\usepackage{booktabs}
\usepackage{hhline}
\usepackage{rotating}
\usepackage{complexity}
\usepackage{color}

\newcommand{\bra}[1]{\left\langle #1 \right|} 
\newcommand{\ket}[1]{\left| #1 \right\rangle}

\DeclareMathOperator{\Tr}{Tr}
\DeclareMathOperator{\Var}{Var}

\begin{document}

\title{Experimental robust self-testing of the state generated by a quantum network}

\author{Iris Agresti}
\affiliation{Dipartimento di Fisica, Sapienza Universit\`{a} di Roma,
Piazzale Aldo Moro 5, I-00185 Roma, Italy}
\author{Beatrice Polacchi}
\affiliation{Dipartimento di Fisica, Sapienza Universit\`{a} di Roma,
Piazzale Aldo Moro 5, I-00185 Roma, Italy}
\author{Davide Poderini}
\affiliation{Dipartimento di Fisica, Sapienza Universit\`{a} di Roma,
Piazzale Aldo Moro 5, I-00185 Roma, Italy}
\author{Emanuele Polino}
\affiliation{Dipartimento di Fisica, Sapienza Universit\`{a} di Roma,
Piazzale Aldo Moro 5, I-00185 Roma, Italy}
\author{Alessia Suprano}
\affiliation{Dipartimento di Fisica, Sapienza Universit\`{a} di Roma,
Piazzale Aldo Moro 5, I-00185 Roma, Italy}
\author{Ivan \v{S}upi\'{c}}
\affiliation{D{\'{e}}partement de Physique Appliqu\'{e}e, Universit\'{e} de Gen\`{e}ve, 1211 Gen\`{e}ve, Switzerland}
\author{Joseph Bowles}
\affiliation{ICFO - Institut de Ciencies Fotoniques, The Barcelona Institute of Science and Technology, E-08860 Castelldefels, Barcelona, Spain}
\author{Gonzalo Carvacho}
\affiliation{Dipartimento di Fisica, Sapienza Universit\`{a} di Roma,
Piazzale Aldo Moro 5, I-00185 Roma, Italy}
\author{Daniel Cavalcanti}
\email{Daniel.Cavalcanti@icfo.eu}
\affiliation{ICFO - Institut de Ciencies Fotoniques, The Barcelona Institute of Science and Technology, E-08860 Castelldefels, Barcelona, Spain}
\author{Fabio Sciarrino}
\email{fabio.sciarrino@uniroma1.it}
\affiliation{Dipartimento di Fisica, Sapienza Universit\`{a} di Roma,
Piazzale Aldo Moro 5, I-00185 Roma, Italy}

\begin{abstract} 

Self-testing is a method of quantum state and measurement estimation that does not rely on assumptions about the inner working of the used devices. Its experimental realization has been limited to sources producing single quantum states so far. 
In this work, we experimentally implement two significant building blocks of a quantum network involving two independent sources, i.e. a parallel configuration in which two parties share two copies of a state, and a tripartite configuration where a central node shares two independent states with peripheral nodes. Then, by extending previous self-testing techniques we provide device-independent lower bounds on the fidelity between the generated states and an ideal state made by the tensor product of two maximally entangled two-qubit states. Given its scalability and versatility, this technique can find application in the certification of larger networks of different topologies, for quantum communication and cryptography tasks and randomness generation protocols.
\end{abstract}

\maketitle

\section{Introduction}
Within the last few years, a large number of quantum resource-based protocols have been designed, with a wide range of applications. However, it is crucial, and far from trivial, to discriminate the devices that are working correctly from those that are not. Indeed, two difficulties can emerge: on one hand, the task required by the user may be hard to verify (a notorious example being the Boson Sampling problem \cite{aaronson_2011, broome_2013, spring_2013, tillmann_2013, crespi_2013}) and, on the other, the devices may be affected by noise and imperfections that are unknown to the user. The latter case is especially relevant for tasks aimed to be secure against eventual adversaries, which could exploit such defects to obtain secret information or sabotage the operation of the devices. For instance, this is the case of private randomness generation or amplification and quantum key distribution protocols \cite{acin_2007, ramanathan_2016_prl, colbeck_2012, gallego_2013, miller_2016, brandao_2016, vazirani_2011, liu_2018, vazirani_2014, chung_2014, dupuis_2016, friedman_2015, shen_2018, kessler_2017, bancal_2014, bierhorst_2018}. Hence, the ability to certify that the device is operating properly, and possibly without relying on the knowledge of its internal working, is crucial for a wider application of quantum technologies. The approach where conclusions about the correctness of the device's operation are drawn only from input/output statistics, is known as \textit{device-independent} (DI) \cite{pironio_focus} and typically relies on the quantum violation of Bell-like inequalities \cite{brunner_aps_2014}.

A key protocol in the device-independent scenario is that of self-testing \cite{STreview}. There, a multipartite quantum state is subject to a number local measurements (a procedure called a Bell test), and the resulting statistics alone are enough to certify the specific form of the state and measurements. For instance, obtaining the maximum value of $2\sqrt{2}$ in a Clauser-Horne-Shimony-Holt (CHSH) Bell test \cite{chsh} certifies that the state is equivalent to a two-qubit maximally entangled state. In recent years several self testing protocols have been proposed to certify different states and measurements \cite{mayers_2003, wang_2016, bamps_2015, supic_2016, kaniewski_2017, yang_2014, kaniewski_2016, coladangelo_2017,  supic_2018, renou_2018,  sekatski_2018, rocchetto_2019, bowles_2018, gomez_2019, breiner_2018, bancal_2015}. In this work we present  experimental demonstrations of self-testing for two types of quantum network, each featuring two independent sources (see Fig.~\ref{fig:scenarios}): (i) a network in which the sources are placed in a parallel configuration between two parties, and (ii) a network featuring three parties, where a central party shares a source with two peripheral parties. The design of our experimental setup follows the bipartite self-testing strategies recently proposed in \cite{supic_2019}, which we further adapt to the multipartite network (ii). To experimentally implement the network structures, we employ a flexible and versatile platform, introduced in \cite{poderini_network}, which allows one to easily change the quantum network topology. Precise lower bounds on the self-testing fidelity with the desired states are obtained from the experimental statistics via the SWAP method \cite{yang_2014,bancal_2015}, a numerical tool based on semidefinite programming.
Our results show that under realistic experimental conditions we can obtain non-trivial device-independent lower bounds on the fidelity between the actual state and ideal states. Moreover, the present techniques can in principle  be extended to an arbitrary number of nodes and to an arbitrary target state, which makes them a promising tool for the certification of larger networks and in the implementation of quantum communication and cryptography tasks.

\section{Self-testing of quantum networks}

\begin{figure}
\includegraphics[width=\columnwidth]{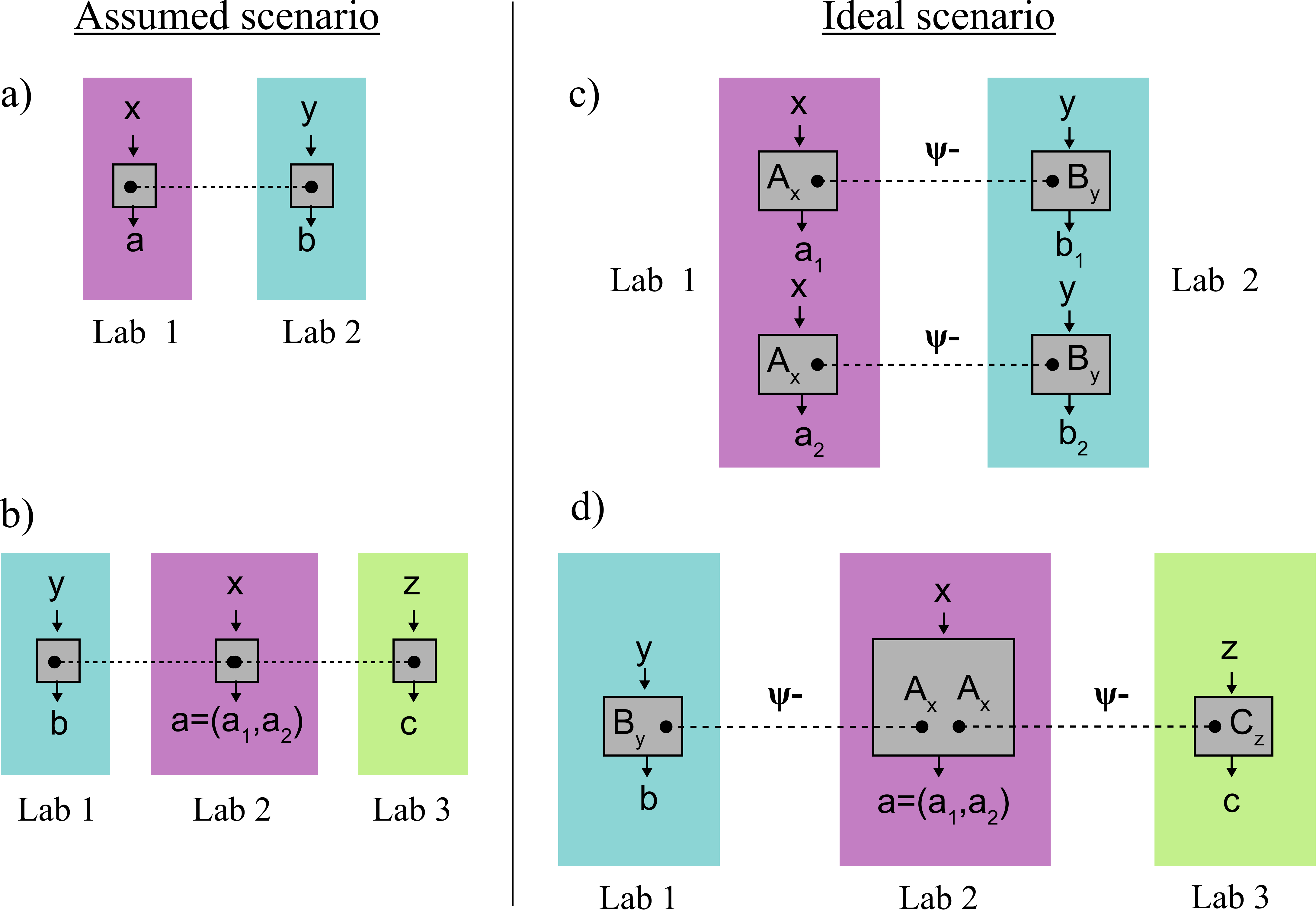}
\caption{\textbf{Self-testing scenarios.} The self-testing procedure consists in performing an experiment and analyse the produced data without assuming a particular implementation (\textbf{a},\textbf{b}), i.e. by considering a black-box scenario in which we only have access to the conditional probability distributions of measurement results (labeled by $a,b,$ and $c$) conditioned to measurement choices (labeled by $x,y$ and $z$). Nothing is assumed on the shared quantum state and measurements. Then self-testing techniques are used to obtain the minimum fidelity between the real states produced in the experiment and the ideal situation shown in the right panel (\textbf{c},\textbf{d}), where the sources produce perfect maximally entangled two-qubit states (e.g. $|\psi^-\rangle$)
and each party applies the local Pauli measurements to each qubit, corresponding to a maximal violation of the CHSH inequality, e.g $A_0=\sigma_z, A_1=\sigma_x,B_0=C_0=-(\sigma_x+\sigma_z)/\sqrt{2}, B_1=C_1= (\sigma_x-\sigma_z)/\sqrt{2}$.
Here we perform self-testing analysis of the state produced in two geometries: \textbf{c)} a bipartite situation where we aim at certifying the presence of two copies of a maximally entangled state and \textbf{d)} a tripartite scenario in which a central party shares maximally entangled states with two peripheral parties.}
\label{fig:scenarios}
\end{figure}

In the device-independent scenario the measurement devices and sources are treated as black boxes, exchanging only classical communication with external users. Suppose the users (which we label as $A,B,C,\ldots$) share some state $\rho_{ABC\cdots}$ that is unknown to them, and that they can prepare and measure the state in an independent, identically distributed  (i.i.d.) manner. After the experiment is repeated many times the users can estimate the probabilities $p(a,b,c,\ldots|x,y,z,\ldots)$ of obtaining measurement outcomes $a,b,c,\ldots,$ if measurements $x,y,z,\ldots$ are performed. According to quantum mechanics, such probabilities are given, through the Born rule as
\begin{align}
    p(a,b,c,\ldots|x,y,z,\ldots)=\Tr[ \rho_{ABC\cdots} \, A_{a}^x \otimes B_{b}^{y} \otimes C_{c}^z\otimes \cdots ],
    \label{eq:probs}
\end{align}
where $A_{a}^x,B_{b}^y,C_{c}^z\dots$ denote the local measurement operators. We say that the probabilities $p(a,b,c,\ldots|x,y,z,\ldots)$ self-test the target state $\ket{\psi'}_{ABC\cdots}$ if the observation of $p(a,b,c,\ldots|x,y,z,\ldots)$ necessarily implies the existence of a local quantum channel $\Omega[\cdot]=\Omega_A [\cdot]\otimes\Omega_B [\cdot]\otimes\Omega_C [\cdot]\cdots$ such that 
\begin{align}\label{perfectST}
    \Omega[\rho_{ABC\cdots}] = \ket{\psi'}\bra{\psi'}_{ABC\cdots}.
\end{align}
Self-testing therefore certifies that the parties share the state $\ket{\psi'}_{ABC\cdots}$, in the sense that there exist local operations the parties could perform to extract the state from $\rho$. Note that this statement holds for \emph{any} state $\rho$ satisfying \eqref{eq:probs} (for some local measurements) and is thus a device-independent statement. As an example, it is known that any bipartite state $\rho_{AB}$ producing correlations that give the maximal quantum violation of the CHSH Bell inequality (with value $2\sqrt{2}$), there exists a channel such that $(\Omega_A\otimes\Omega_B)[\rho_{AB}]=\ket{\psi^-}\bra{\psi^-}$, with $\ket{\psi^-}=\frac{1}{\sqrt{2}}(\ket{01}-\ket{10})$ the maximally entangled singlet state. 

In realistic scenarios, however, it is impossible to exactly meet the self-testing conditions \eqref{eq:probs}, not only due to experimental noise, but also because the finite experiment time implies that one can only infer the probabilities up to a given confidence interval. For this reason, the self-testing statement has to be robust, i.e. to be able to say something about the underlying state even when the self-testing condition is only approximately met. To do this, we focus on lower bounding the fidelity between the extracted state and the target state
\begin{align}
    F(\Omega[\rho_{ABC\cdots}],\ket{\psi'}\bra{\psi'}_{ABC\cdots})=\bra{\psi'}\Omega[\rho_{ABC\cdots}]\ket{\psi'}
    \label{eq:fidelity}
\end{align}
given the experimental statistics. That is, one proves that for any state producing the experimental statistics there exists a local channel $\Omega$ such that $F\geq f$ (for some $f\leq 1$), up to a given confidence level. Note $f=1$ corresponds to the case of perfect self-testing given in \eqref{perfectST}. A useful method that we use for calculating such lower bounds is the SWAP method \cite{yang_2014,bancal_2015}, a numerical tool based on semidefinite programming and the Navascu\'{e}s-Pironio-Ac\'{i}n (NPA) hierarchy. In particular, this technique amounts to numerically swapping part of the generated state on a dummy register, in order to find a proper expression for the fidelity \eqref{eq:fidelity}, as a function of the correlations terms $p(a,b,c \dots \vert x,y,z \dots)$. Then, a lower bound on this fidelity can be obtained through an SDP, over a superset of the quantum correlations set, mathematically defined by a level $l$ of the NPA hierarchy. In order to get tighter bounds, further linear constraints can be added to the problem, e.g. the observed correlations $p(a,b,c \dots \vert x,y,z \dots)$. More details about this method can be found in the Supplementary Information.  

In this work we report on the self-test of target states that correspond to two independent sources producing maximally entangled singlet states $\ket{\psi^-}$. We focus on two scenarios, depicted in Fig.~\ref{fig:scenarios}, which can be seen as two possible building blocks of a quantum network. The first scenario we consider features two parties ($A$ and $B$), and the target state we self-test corresponds to preparing the two maximally entangled states in parallel (see Fig.~\ref{fig:scenarios}c). That is, we self-test the state $|\Psi \rangle_2 = |\psi^- \rangle_{A_{1}B_{1}} \otimes |\psi^-\rangle_{A_{2}B_{2}}$,  where $A_{i}$, $B_j$ denote local qubit Hilbert spaces of party $A,B$. The second structure we consider features three parties ($A,B,C$), and the target state corresponds to preparing the sources in an entanglement swapping network (see Fig.~\ref{fig:scenarios}d). This configuration, although it has been investigated, both theoretically and experimentally, within the last years \cite{branciard_2010, branciard_2012, Carvacho2017, saunders_2017, andreoli_2017}, was only very recently implemented exploiting truly independent sources \cite{sun_2019, poderini_network} and closing the locality loophole \cite{sun_2019}. Our target state in this scenario is therefore $|\Psi\rangle_3 = |\psi^- \rangle_{A_{1}, B} \otimes |\psi^- \rangle_{A_{2}, C}$. We stress that, although the target states correspond to states produced by two independent sources, we do not assume this structure at the black-box level, that is, the parties can in principle share any multipartite state, but the self-testing statements ensure they have the desired product form. 

Our self-testing protocol is inspired from the recent work \cite{supic_2019}, which presents a method for self-testing tensor products of copies of a state, while keeping the number of inputs constant. Such a method is desirable for self-testing quantum networks, since standard methods feature a number of inputs that grows exponentially with the number of copies, which will become a practical issue for larger networks. We consider scenarios in which all parties have two inputs; $x,y=0,1$ for scenario (a) and $x,y,z=0,1$ for scenario (b). The number of outputs is given by the local Hilbert space dimension of the target state: in scenario (a) both parties have four outputs, which we write as $a=(a_1,a_2)$ and $b=(b_1,b_2)$, where $a_i$, $b_i$ take values $\pm1$; in scenario (b) we have $a=(a_1,a_2)$ as before and $b=\pm1$, $c=\pm1$. The measurements are chosen so that in the ideal experiment, the marginal distributions maximally violate the CHSH Bell inequality. More precisely one has $\text{CHSH}(p(a_i,b_i\vert x,y))=2\sqrt{2}$, $i=1,2$ for scenario (a) and $\text{CHSH}(p(a_1,b\vert x,y))=\text{CHSH}(p(a_2,c\vert x,z))=2\sqrt{2}$ for scenario (b). Following results from \cite{supic_2019} such distributions are known to self-test the desired target states. The measurement strategy corresponds to the performing the standard CHSH measurements in parallel, which we elaborate on in the following section. More details about merging the SWAP method with the techniques from \cite{supic_2019} can be found in the Supplementary Information.

\begin{figure*}[t!]
\includegraphics[width=\textwidth]{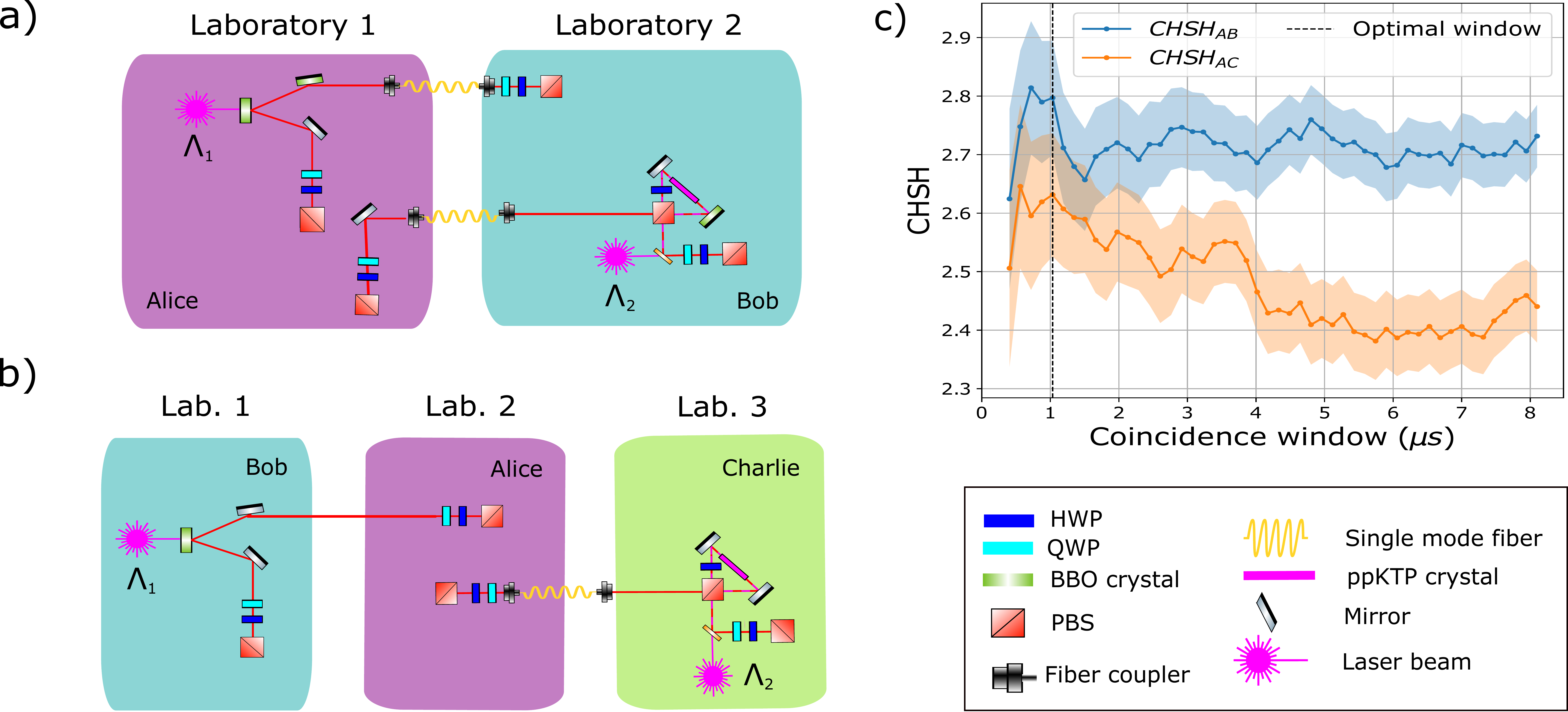}
\caption{\textbf{Experimental apparatus and detection of four-fold coincidences.} \textbf{a)} Parallel self-testing scenario implementation. In this scenario, the involved laboratories are two, both equipped with a quantum state source and two measurement stations. The measurement stations in Laboratory 1 represent Alice, while those in Laboratory 2, Bob. Both the sources generate a bipartite entangled state, send a subsystem to the other laboratory through a $\sim 30~m$ long single-mode fiber and keep the other one to measure it. The target state to be shared between the two parties involved in this network is the tensor product of two maximally entangled two-qubit states.  \textbf{b)} Tripartite scenario implementation. In this scenario, the involved laboratories are three, two (1 and 3) equipped with quantum state source and a measurement station (representing, respectively, Bob and Charlie), and one, just with 2 measurement stations, constituting Alice. The source in laboratory 1, $\Lambda_{1}$, sends one photon to Bob and one to Alice, while $\Lambda_{2}$ sends one to Charlie and the other one to Alice, through a $\sim 30~m$ long single-mode fiber. \textbf{c)} In order to detect the couples of 2-qubit states generated by sources $\Lambda_1$ and $\Lambda_2$, we set a coincidence window within which two two-fold coincidence events must occur, to be recognized as a four-fold coincidence. The curves indicate the value of the CHSH of the states generated by the two sources ($\Lambda_1$ blue curve, $\Lambda_2$ orange curve), with the statistical uncertainty obtained considering the Poissonian distribution of the events (1 standard deviation), in terms of such a 4-fold coincidences window. The optimal time interval for this window is then chosen by evaluating the weighted mean of the two violations and choosing the highest one. In this case, it amounts to $1.033~ \mu s$.}
\label{fig:expw}
\end{figure*}

\section{Experimental apparatus}

In the experimental implementation of the two scenarios of interest, we resort to the versatile photonic platform introduced in \cite{poderini_network}. In particular, for the parallel self-testing case (Fig.~\ref{fig:scenarios}c) we employ two separate laboratories equipped with independent quantum state sources (respectively, $\Lambda_1$ and $\Lambda_2$) and two measurement stations. Each measurement station is composed by a half-wave plate (HWP) and a polarizing beam splitter (PBS), which allows to perform polarization projective measurements of the form $\cos({4\theta})\sigma_{z}+ \sin({4\theta})\sigma_{x}$, where $\sigma_{x}$ and $\sigma_{z}$ are Pauli operators, by simply rotating the HWP of the angle $\theta$ with respect to its optical axis. In the end, all the registered counts are sent to a central counter, which recognizes as coincidence events of distant detectors, the counts within a given time window. In our notation, the measurement stations in laboratory $1$ represent Alice, while those in laboratory $2$, Bob. Then, the two laboratories are connected through two $\sim 30~m$ long single-mode fibers, as shown in Fig.~\ref{fig:expw}a. The source in laboratory $1$ employs spontaneous parametric down-conversion (SPDC) of type II to generate a pair of polarization entangled photons of wavelength equal to $\lambda=785~nm$, through a Beta-Barium Borate (BBO) crystal which is pumped, in pulsed regime, by a $\lambda = 392.5 nm$ $UV$ laser beam. Instead, in laboratory $2$, we have a periodically polarized Tytanil Phosphate, pumped in the continuous-wave regime, which generates polarization entangled photon pairs at $\lambda=808~nm$. 
Both sources generate a 2-qubit maximally entangled state, e.g. the singlet state $|\psi^{-}\rangle$, where the computational basis ($\ket{0}$ and $\ket{1}$) is encoded in the horizontal and vertical photon polarization states ($\ket{H}$ and $\ket{V}$), hence $\ket{\psi^{-}}=\frac{\ket{HV}-\ket{VH}}{\sqrt{2}}$. At this point, both laboratories send a photon to the other one and use one measurement station to perform projective measurements on the photon they kept and the second one to measure the photon they received. In detail, Alice's and Bob's operators will be those maximising the CHSH inequality violation, i.e., up to unitary transformations, $A_{0}=A_{0}^1 \otimes A_{0}^2=\sigma_z \otimes \sigma_z$ and $A_{1}=A_{1}^1 \otimes A_{1}^2=\sigma_x \otimes \sigma_x$ and $B_{0}=B_{0}^1 \otimes B_{0}^2=\frac{(\sigma_x + \sigma_z)}{\sqrt{2}} \otimes \frac{(\sigma_x + \sigma_z)}{\sqrt{2}}$ and $B_{1}=B_{1}^1 \otimes B_{1}^2=\frac{(\sigma_x - \sigma_z)}{\sqrt{2}} \otimes \frac{(\sigma_x - \sigma_z)}{\sqrt{2}}$, indicating with the superscript the source generating the subsystem.

In order to detect coincidence events between distant detectors, we designed a software to coordinate the counters located in the different laboratories. In particular, the software considered two coincidence time windows: one to detect the 2-fold coincidences generated by each source (set to $1.05~ns$), $w_1$, and another to reveal 4-fold coincidences, i.e. simultaneous 2-fold events occurring for both sources, $w_2$. In other words, if a 2-fold event is registered for both sources, within $w_2$, those are labelled as simultaneous and considered as a 4-fold event.  Then, the optimal value for $w_2$ is chosen through the corresponding CHSH values brought by the two sources, as depicted in Fig.~\ref{fig:expw}c, and choosing the one giving the highest weighted average between the two values.

In the tripartite scenario (Fig.~\ref{fig:scenarios}d) we have 3 laboratories. Laboratories $1$ and $3$ (the peripherical nodes) are constituted by a quantum state source and one measurement station each, while laboratory $2$ (the central node) has two measurement stations, as shown in Fig.~\ref{fig:expw}b. The source in laboratory $1$ sends one photon to Bob's measurement station and the other to Alice, while the source in laboratory $3$ will send one photon to Alice the other one to Charlie. In this case, analogously to before, the measurement operators will be the following: $A_{0}=A_{0}^1 \otimes A_{0}^2=\sigma_z \otimes \sigma_z$ and $A_{1}=A_{1}^1 \otimes A_{1}^2=\sigma_x \otimes \sigma_x$; $B_{0}=-\frac{(\sigma_x + \sigma_z)}{\sqrt{2}}$ and $B_{1}=\frac{(\sigma_x - \sigma_z)}{\sqrt{2}}$;  $C_{0}=-\frac{(\sigma_x + \sigma_z)}{\sqrt{2}}$ and $C_{1}=\frac{(\sigma_x - \sigma_z)}{\sqrt{2}}$.

\section{Results}

Our main goal is to experimentally obtain lower bounds on the fidelities between the produced states and the reference states, based on the statistics observed through the two apparatuses shown in Fig.~\ref{fig:expw}a-b. However, we cannot simply apply the SWAP method using the raw statistics, because due to finite statistics the experimental frequencies do not correspond to physically allowed correlations (for instance, they violate the no-signalling conditions) and the optimization constraints imposed by the NPA would make the problem infeasible. Therefore, to overcome this problem, we proceed in the following way (see Supplementary Information for more details):

\begin{enumerate}

\item We use a regularization method in which we approximate the experimental frequencies $f_j$ by probability distributions belonging to the NPA set $\mathcal{Q}_4$ \cite{lin2018device}, where $j=(a,b,x,y)$ or $j=(a,b,c,x,y,z)$ depending on the scenario. The solution of this method provides a no-signalling set of distributions $P_j^{reg}$ that are guaranteed to be close to the set of quantum distributions. See Methods for details. 

\item We then run the SWAP SDP using $P_j^{reg}$ as inputs. 

The solution of this SDP provides a linear functional $d(P)=\sum_j c^*_j P_j$ (through its dual formulation) for which the value gives a lower bound on the self-testing fidelity.

\item We run another SWAP SDP in which we do not assume the actual value of the distributions, but we impose as a constraint the experimentally obtained value of the dual functional.
\end{enumerate}

\begin{figure*}[ht!]
    \centering
    \includegraphics[width=1\textwidth]{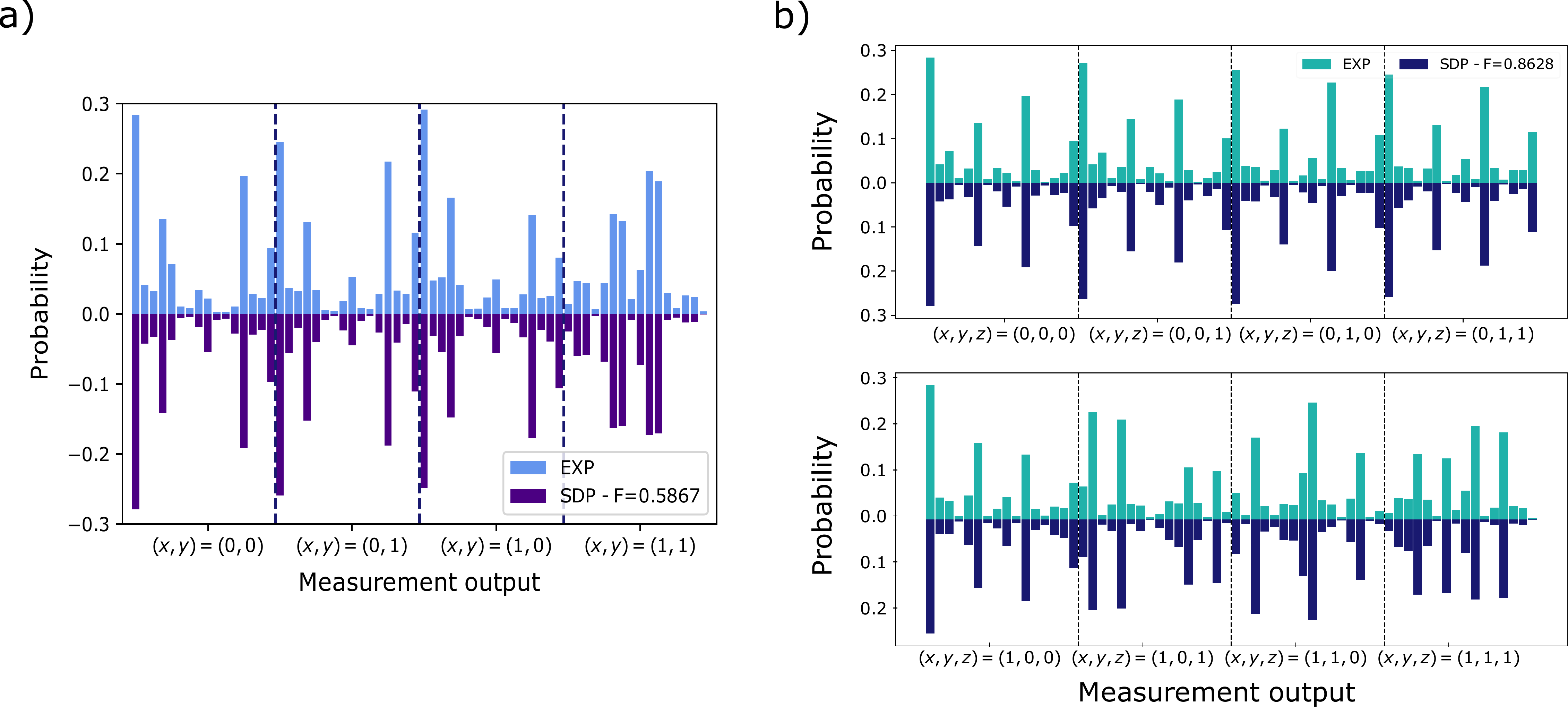}
    \caption{\textbf{Certifiable fidelities in parallel self-testing and tripartite scenarios (Dual method).} In the shown histograms, the upper bars correspond to the experimental frequencies of each measurement output, while the lower ones to the probabilities given as solution by the SDP optimization, with the experimental dual inequality imposed as a constraint. On the x axis, frequencies/probabilities are ordered in blocks that correspond to the possible operators' choices. \textbf{a)} Parallel self testing scenario: indigo columns represent the experimental probabilities, divided in blocks that correspond to eight different sets of operators $(x,y)$ respectively for Alice and Bob. Every block contains sixteen columns, each corresponding to a different set of outcomes: $(a,b)=(0,0), (0,1), (0,2), (0,3), (1,0), ...$, with $a=0,1,2,3$ and $b=0,1,2,3$. Purple columns represent the probabilities found by the SDP program, with the experimental dual inequality imposed as a constraint, and corresponding to the computed bound for the fidelity $F=0.5867$.
     \textbf{b)} Three parties case: turquoise columns represent the experimental probabilities obtained by gathering all our data sockets, divided in blocks that correspond to eight different sets of operators $(x,y,z)$ respectively for Alice, Bob and Charlie. Every block contains sixteen columns, each corresponding to a different set of outcomes: $(a,b,c)=(0,0,0), (0,0,1), (0,1,0), (0,1,1), (1,0,0), ...$, with $a=0,1,2,3$ and $b,c=0,1$. Blue columns represent the probabilities found by the SDP program, with the experimental dual inequality imposed as a constraint, and corresponding to the computed bound for the fidelity $F=0.8628$.}
    \label{fig:hist}
\end{figure*}

Using this method and a Monte Carlo simulation (assuming Poissonian statistics for the experiment) to calculate the uncertainties, we found that $F(\rho_{AB},\ket{\Psi}_2)=0.587 \pm 0.053$ for the parallel configurations and $F(\rho_{ABC },\ket{\Psi}_3)=0.863 \pm 0.032$ for the tripartite case. In detail, in Fig.\ref{fig:hist}, we show the observable terms of the probability distribution given as solution by the SDP run in the third step, in comparison to the experimental frequencies.

\subsubsection{Device-Independent estimation of the uncertainty}

The previous method for calculating experimental uncertainties are not fully device-independent as it assumes a Poissonian distribution for the measurement results. We now move a step forward in removing assumptions and quantify the confidence level on the fidelities bounds exploiting Hoeffding's inequality \cite{hoeffding_1963}, which holds for independent variables that are in the range $(0,1)$. 

\begin{figure}[ht!]
    \centering
    \includegraphics[width=1\columnwidth]{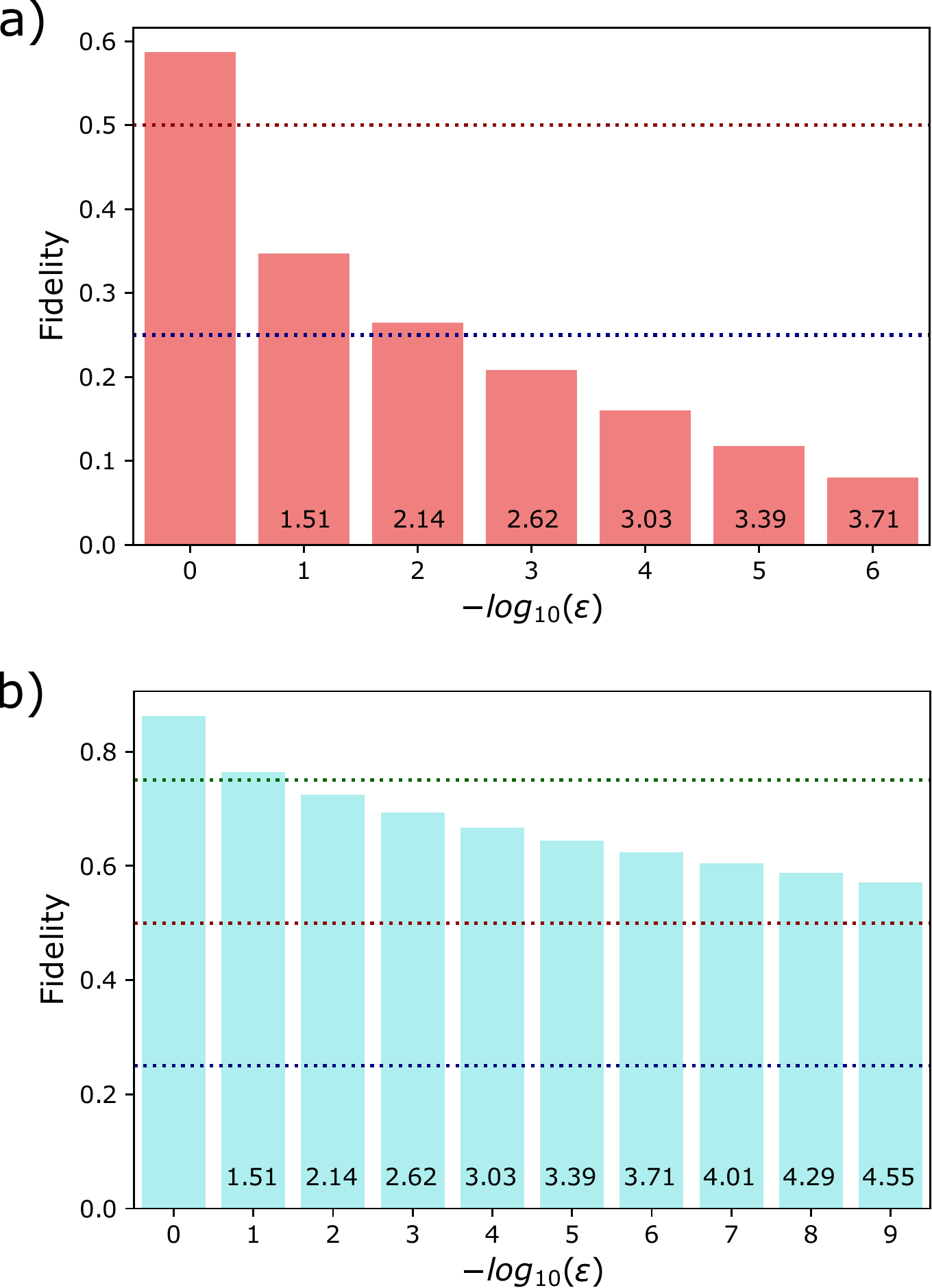}
    \caption{\textbf{Certifiable fidelities and Hoeffding's confidence levels.} We show the lower bounds on the certifiable fidelities, in both of the studied scenarios, as a function of the probability that for all configurations $j$, $f_j \notin \{P_j - t_j(\epsilon), P_j +t_j(\epsilon) \}$, where $f_j$ are the observed frequencies, $P_j$ the probabilities characterizing the probability distribution underlying the experiment and $n_j$ the registered counts corresponding to configuration $j$. Furthermore, by Hoeffding's inequality \cite{hoeffding_1963}, $t_{j}(\epsilon)=\sqrt{\frac{-\ln(\epsilon)}{2n_j}}$. From those statistical uncertainties on the probabilities, we recovered the confidence level on the obtained lower bounds, which amounts to $\sqrt{-\ln(\epsilon)}$ standard deviations of a Gaussian distribution (see Supplementary Information for a full derivation). In \textbf{a)}, we report the case of parallel self-testing, while, in \textbf{b)}, the tripartite one. The numbers on the bars indicate the number of standard deviations corresponding to the confidence level of such fidelities. Such confidence levels, then, can be found on a standard normal table. The dashed lines indicate the fidelity values that certify, respectively, that the state has a Schmidt number higher or equal to 2 (blue line), 3 (red line) and 4 (green line).}
    \label{hoeff}
\end{figure}

For this second method, we relax the constraint that we obtain a given value for the SDP functional $d(P)$ and only assume bound to it given by $d(P) \leq d^{exp} + \tau (\epsilon)$ and $d(P) \geq d^{exp} - \tau (\epsilon)$. In this notation, $1-\epsilon$ constitutes the confidence level that the observed frequency $f_j$ are within a range $t_j(\epsilon)$ from the real probability $P_j$ \cite{hoeffding_1963}. 
More specifically, such interval $t_j(\epsilon)$ amounts to $\sqrt{\frac{-\ln(\epsilon)}{2 n_{j}}}$, where $n_j$ is the number of registered counts for configuration $j$. At this point, by the central limit theorem \cite{bini}, the linear combination of frequencies $d^{exp}$ will be characterized by a Gaussian statistics, whose variance amounts to $\sigma^2= \sum_{j} {c^*}_{j}^2 Var(f_j)$. Furthermore, given that $\Var(f_j)=\frac{1}{2n_j}$, parameter $\tau(\epsilon)$ was chosen as follows (see the Supplementary Information for full derivation):
\begin{equation}
    \tau(\epsilon)^2=\sum_{j} {c^*}_{j}^2 t(\epsilon)^2_{j}= -\ln(\epsilon) \sigma^2
\end{equation}
In the end, the confidence level that the true value of $d(P)$ is within a range of $\tau(\epsilon)$ from $d^{exp}$ can be easily recovered from a standard normal table, considering that this interval amounts to $\sqrt{-\ln(\epsilon)}$ standard deviations.
In Fig.~\ref{hoeff}, we show the certifiable fidelities in the two studied scenarios, versus $\epsilon$ and indicate the corresponding number of standard deviations adding up to $\tau(\epsilon)$ at the bottom of the bars. From such fidelities, we can extrapolate a lower bound on the Schmidt number of the state \cite{gupta_2015, zeilinger_interface}. In particular, in the studied cases, a fidelity higher than $\frac{k}{4}$ implies a Schmidt number (SN) higher or equal to $k+1$. In Fig.~\ref{hoeff}, we indicate the thresholds for $SN=2$ (blue line), $3$ (red line) and $4$ (green line).

\section{Discussion}

In this work, we experimentally implemented device-independent self-testing protocols in two scenario representing two basic building blocks for quantum networks. In particular, we studied a parallel self-testing scenario, in which two parties share two copies of a bipartite state, and a tripartite one, in which two bipartite states are shared among two peripheral nodes and a central one \cite{Carvacho2017, andreoli_2017, saunders_2017, sun_2019, branciard_2010, branciard_2012}. 
In these two cases, we were able to obtain lower bounds on the fidelity of the generated states with the desired target states that demonstrate that both sources indeed produce entangled states. In detail, we found lower bounds on the Schmidt number of the generated states, certifying a $SN \geq 2$ for the first case and a $SN \geq 3$ for the second one.

We stress that we considered the scenario presented in Ref. \cite{supic_2019} where the number of local measurement choices are kept constant independently of the number of quantum state copies that one aims at certifying. Because of this, in our implementation each party had to implement only two measurements basis each, although the self-testing protocol considered two copies of maximally entangled states. From an experimental perspective, this method provides an extra significant advantage, represented by the fact that this technique requires only separable measurements. In particular, this result shows that, with a simple platform and low resource expense, it is possible to obtain non-trivial device-independent estimates of the quality of quantum networks. We believe that the present tool will find future applications in quantum communication, in particular in cryptographic scenarios such as quantum key distribution and blind quantum computation.

\section{Methods}
\subsection*{Experimental details}

For the experimental setups of Fig.~\ref{fig:expw}, the pump laser for source
$1$, with $\lambda=392.5$~nm are produced by a second harmonic generation
(SHG) process from a Ti:Sapphire mode locked laser with repetition rate of $76$ MHz. Photon 
pairs entangled in the polarization degree of freedom are generated exploiting type-II SPDC in 2
mm-thick beta-barium borate (BBO) crystals.
Source $2$, instead, employs a continuous wave diode laser with wavelength of 
$\lambda=404$~nm, which pumps a 20mm-thick periodically-poled KTP crystal inside a Sagnac interferometer, 
to generate photon pairs using a type-II degenerate SPDC process.
The photons generated in all the sources are filtered in wavelength and spatial mode by using narrow band interference filters and single-mode fibers, respectively.

\subsection*{Coincidence counting}
The photon detection events were collected and timed by a different time tagger device for each party, located in the corresponding laboratory \cite{poderini_network}.
For each $1$~s of data acquisition the events were sent to a central server, along with a random clock signal shared between all the time-taggers, which was used to
synchronize the timestamps of events relative to different devices.
To filter out part of the noise the raw data was first pre-processed by keeping only double coincidence events for each photon source, using a narrow coincidence window of $1.05$~ns.
Then, the 4-fold coincidence events between the two sources were counted every time one of such double coincidence event was recorded for each source in a window of $\sim 1.033~ \mu s$.

\section*{Acknowledgements} 

This work was supported by The John Templeton Foundation via the grant Q-CAUSAL No 61084, by MIUR (Ministero dell'Istruzione, dell'Università e della Ricerca), via project PRIN 2017 "Taming complexity via QUantum Strategies a Hybrid Integrated Photonic approach" (QUSHIP) Id. 2017SRNBRK, by the Regione Lazio programme “Progetti di Gruppi di ricerca” legge Regionale n. 13/2008 (SINFONIA project, prot. n. 85-2017-15200) via LazioInnova spa. I.A. acknowledges La Sapienza University of Rome via the grant for joint research projects for the mobility n.~2289/2018 Prot. n.~50074. DC acknowledges support from the Government of Spain (Ramon y Cajal fellowship, FIS2020-TRANQI and Severo Ochoa CEX2019-000910-S), Fundació Cellex, Fundació Mir-Puig, Generalitat de Catalunya (CERCA, AGAUR SGR 1381), ERC AdG CERQUTE and Swiss National Science Foundation (Starting Grant DIAQ and NCCR QSIT).

\end{document}


\title{Supplementary Information\\ Experimental robust self-testing of the state generated by a quantum network}

\author{Iris Agresti}
\affiliation{Dipartimento di Fisica, Sapienza Universit\`{a} di Roma,
Piazzale Aldo Moro 5, I-00185 Roma, Italy}
\author{Beatrice Polacchi}
\affiliation{Dipartimento di Fisica, Sapienza Universit\`{a} di Roma,
Piazzale Aldo Moro 5, I-00185 Roma, Italy}
\author{Davide Poderini}
\affiliation{Dipartimento di Fisica, Sapienza Universit\`{a} di Roma,
Piazzale Aldo Moro 5, I-00185 Roma, Italy}
\author{Emanuele Polino}
\affiliation{Dipartimento di Fisica, Sapienza Universit\`{a} di Roma,
Piazzale Aldo Moro 5, I-00185 Roma, Italy}
\author{Alessia Suprano}
\affiliation{Dipartimento di Fisica, Sapienza Universit\`{a} di Roma,
Piazzale Aldo Moro 5, I-00185 Roma, Italy}
\author{Ivan \v{S}upi\'{c}}
\affiliation{D{\'{e}}partement de Physique Appliqu\'{e}e, Universit\'{e} de Gen\`{e}ve, 1211 Gen\`{e}ve, Switzerland}
\author{Joseph Bowles}
\affiliation{ICFO - Institut de Ciencies Fotoniques, The Barcelona Institute of Science and Technology, E-08860 Castelldefels, Barcelona, Spain}
\author{Gonzalo Carvacho}
\affiliation{Dipartimento di Fisica, Sapienza Universit\`{a} di Roma,
Piazzale Aldo Moro 5, I-00185 Roma, Italy}
\author{Daniel Cavalcanti}
\email{Daniel.Cavalcanti@icfo.eu}
\affiliation{ICFO - Institut de Ciencies Fotoniques, The Barcelona Institute of Science and Technology, E-08860 Castelldefels, Barcelona, Spain}
\author{Fabio Sciarrino}
\email{fabio.sciarrino@uniroma1.it}
\affiliation{Dipartimento di Fisica, Sapienza Universit\`{a} di Roma,
Piazzale Aldo Moro 5, I-00185 Roma, Italy}

\maketitle

\section{Bound on the fidelity of the state of a quantum network with a target state}
In this section, we describe in further detail the \textit{self-testing} approach we adopted in this work, firstly introduced in \cite{yang_2014}, which can be applied when experimental imperfections do not allow to reach the maximal quantum violation of a causal constraint, like the CHSH inequality. Indeed, in this case, the violation reveals the presence of non-classical correlations, but it does not single out the system producing it. However, through the techniques introduced in \cite{yang_2014}, it is possible to establish a lower bound on the fidelity between the target state and the unknown generated quantum state, resorting to the Navascués-Pironio-Ac\'{i}n (NPA) hierarchy \cite{NPA1, NPA2}.

Indeed, let us consider the simplest quantum scenario, with one source of bipartite entangled system and two parties (Alice and Bob) performing 2-output local measurements on a given subsystem, according to an input $(x,y) \in (0,1)$.
The figure of merit for a \textit{self-testing} of a two-qubit maximally entangled state is the CHSH inequality, but, if the maximal violation extent is not achieved, for instance due to experimental imperfections, the test is inconclusive.

Hence, we can resort to the following protocol: first of all, we use, as figure of merit, the fidelity with a target state $\ket{\psi_{target}}$, defined as follows:
\begin{equation}
    F(\rho_{AB}, \ket{\psi_{target}})=\bra{\psi_{target}} \rho_{AB} \ket{\psi_{target}}
    \label{fidelity_AB}
\end{equation}
However, without making assumptions on the dimension of the generated state, this fidelity is not properly defined, therefore, to have a fully device-independent protocol, we can resort to the SWAP operator \cite{yang_2014}. As a first step, we consider an ancillary register, defined on an Hilbert space of the same dimension than that of the target, and we trust that each system is prepared in the dummy state, i.e. $\ket{0}$. In our case we take $dim(\mathcal{H}_{target})=2$. Then, we define the following local operator 
$S=S_{AA'}\otimes S_{BB'} $, where $S_{AA'}=U_{AA'}V_{AA'}$ and 
\begin{equation}
    U_{AA'}= \mathbb{I} \otimes \ket{0}_{A'}\bra{0}_{A'}+ O_{1}^{A} \otimes \ket{1}_{A'}\bra{1}_{A'}
    \label{u}
\end{equation}
and 
\begin{equation}
    V_{AA'}= \frac{\mathbb{I}+O_{0}^{A}}{2} \otimes \mathbb{I}^{A'} + \frac{\mathbb{I}-O_{0}^{A}}{2} \otimes \sigma_{x}^{A'}
    \label{v}
\end{equation}
and analogously for Bob ($B$). These operations are unitary if both $O_{0}$ and $O_{1}$ are unitary and Hermitian.
Through this operator, we aim to swap part of the state $\rho_{AB}$, which is seen as a black box, onto the ancillary register, and $\rho_{swap}$ will have the following form: 
\begin{equation}
    \rho_{swap}= Tr_{AB}[S(\rho_{AB}\otimes \ket{00}\bra{00})S^{\dag}] 
    \label{partial_trace}
\end{equation}
and, once that we have the $S$ explicitly in terms of $O^{A,B}_{0}$ and $O^{A,B}_{1}$, the entries of $\rho_{swap}$, from the partial trace in Eq.~\eqref{partial_trace}, are given by linear combinations of correlation terms from the set $c=({c_{\mathbb{I}}}=Tr(\rho_{AB}\mathbb{I})$, $c_{O^{A}_{0}}=Tr(\rho_{AB}O^{A}_{0})$, ...~, $c_{O^{A}_{0}O^{A}_{1}O^{B}_{0}}=Tr(\rho_{AB}O^{A}_{0}O^{A}_{1}O^{B}_{0}))$.

Hence, we can solve the following semidefinite program (SDP):
\begin{equation}
    \begin{split}
    f=& \min~\bra{\psi_{target}}\rho_{swap}\ket{\psi_{target}}\\
   & \textrm{s.t.}~c \in \mathcal{Q}_l,\\
    &c_{O_{0}^{A}O^{B}_{0}}+c_{O_{0}^{A}O^{B}_{0}}+c_{O_{0}^{A}O^{B}_{0}}-c_{O_{0}^{A}O^{B}_{0}}=\mathcal{I}_{CHSH}
    \end{split}
\label{fidelity-sdp}
\end{equation}
where $\mathcal{Q}_{l}$ is a set which includes the one of quantum correlations and which corresponds to the $l-th$ level of the NPA hierarchy \cite{NPA1,NPA2}. In this way, by simply evaluating the CHSH inequality on the generated state, and putting it as a constraint in problem \eqref{fidelity-sdp}, we can lower bound the fidelity with the target state. To obtain higher bounds, we can add further constraints, for instance if the statistics corresponds to isotropic black boxes, we could add constraints of the following kind:
\begin{equation*}
    c_{O_{0}^{A}O^{B}_{0}}=c_{O_{1}^{A}O^{B}_{0}}=c_{O_{0}^{A}O^{B}_{1}}=-c_{O_{1}^{A}O^{B}_{1}}
\end{equation*}
A better lower bound can be obtained giving the full statistics to the SDP program, solving:
\begin{equation}
    \begin{split}
    \textrm{given} &~p(a,b|x,y)\\
    f=& \min~\bra{\psi_{target}}\rho_{swap}\ket{\psi_{target}}\\
   & \textrm{s.t.}~c \in \mathcal{Q}_n
    \end{split}
    \label{npa_fidelity}
\end{equation}
The bound on the certifiable fidelity, considering a generated state $\rho_{AB}= v|\psi\rangle \langle \psi| +(1-v) \mathbb{I}/2$ is plotted in Fig.~\ref{fig:fidelities} as a function of the visibility $v$ (dotted curve).
This method is general and can be extended to more parties and different topologies and the operators $O_{0}$ and $O_{1}$ must be chosen according to the target state, although general methods have been introduced, for the case in which one simply has some distributions $p(a,b|x,y)$ and wishes to guess the state and measurements involved \cite{bancal_2015}. 

\subsection{The SWAP method for parallel self-testing}
In the first part of our experiment, we use this protocol for a scenario involving two independent sources $1$ and $2$ generating singlets and two parties performing 4-output local measurements on the subsystem they get, according to an input $(x,y) \in (0,1)$.
Both observers own dummy states with total dimension equal to the one of the target Hilbert space, which is $dim(\mathcal{H}_{target})=4$, so Alice will have $\ket{00}_{A'_1A'_2}$, similarly to Bob. 
The fidelity to be bounded takes the form:
\begin{equation}
    F(\rho_{A_1B_1B_2A_2}, \ket{\psi_{target}})=\bra{\psi_{target}} \rho_{A_1B_1B_2A_2} \ket{\psi_{target}}
    \label{fidelity2parties}
\end{equation}
where the target state is $\ket{\psi_{target}} = \ket{\bar{\psi}} \otimes \ket{\bar{\psi}}$ and $\ket{\bar{\psi}}$ is defined as follows:
\begin{equation}
    \ket{\bar{\psi}} = \cos{\frac{\pi}{8}}\ket{\phi^-} + \sin{\frac{\pi}{8}}\ket{\psi^+}
    \label{psi_bar}
\end{equation}
which is maximally entangled and therefore equivalent to $\ket{\psi^-}$ up to local unitaries. This is chosen for simplicity of notation since this state reaches $ \mathcal{I}_{CHSH} = 2\sqrt{2}$ for the operators:
\begin{equation}
    O^{A}_{0} = \overline{A_{0}} = \overline{B_{0}} = \sigma_z, \quad O^{A}_{1} = \overline{A_{1}} = \overline{B_{1}} = \sigma_x
\end{equation}
to be inserted in the expression for the operators $U$ and $V$ defined in Eq.~\ref{u} and in Eq.~\ref{v}, where $\sigma_{x,z}$ are Pauli operators.
Alice and Bob then perform a swap between the dummy states on their ancillary register and the subsystem they receive, and the $\rho_{swap}$ density matrix takes the following form:
\begin{equation}
    \rho_{swap}= Tr_{A_1B_1A_2B_2}[S(\rho_{A_1B_1A_2B_2}\otimes \ket{0000}\bra{0000})S^{\dag}] 
\end{equation}
where the SWAP operator is defined as $S=S_{A_1A'_1}\otimes S_{B_1B'_1} \otimes S_{A_2A'_2}\otimes S_{B_2B'_2}$. Let us write the complete expression for the SWAP operator acted by Alice $S_{AA'} = S_{A_1A'_1} \otimes S_{A_2A'_2}$, that is analogous for Bob:
\begin{equation}
    S_{AA'} = S_{A_1A'_1} \otimes S_{A_2A'_2} = U_{A_1A'_1A_2A'_2}V_{A_1A'_1A_2A'_2}
\end{equation}
Since the operators $U$ and $V$ act on a subsystem of dimension $2$, Alice and Bob can get $4$ possible outcomes from their measurements, and this can be described by defining $A^{AB}_1 = \sum_{\{x,y\} \in \{0,1\} \times \{0,1\}} (-1)^{Ax+By} \Pi^{2x+y}_{A_1}$, where $\{A,B\}\in\{0,1\}$ and $\Pi_{A_{x}}^{a}$ is the projector on the eigenspace corresponding to the eigenvalue $(a=0,1,2,3)$ of operator $A_{x}$.

With these definitions, the expression of $U$ and $V$ operators has the following form:
\begin{equation}
\begin{split}
U_{A_{1,2},A'_{1,2}}=& A_{1}^{00}\otimes |00\rangle \langle 00| +
 A_{1}^{01}\otimes|01\rangle\langle01|+ \\  &A_{1}^{10}\otimes|10\rangle\langle10|+   A_{1}^{11}\otimes |11\rangle\langle11|
 \end{split}
\label{u_4outcomes}
\end{equation}
\begin{equation}
\begin{split}
V_{A_{1,2},A'_{1,2}}=&\Pi_{A_{0}}^{0} \otimes \mathbb{I} + \Pi_{A_{0}}^{1} \otimes (\mathbb{I} \otimes \sigma_{x}) +\\ &\Pi_{A_{0}}^{2} \otimes (\sigma_{x} \otimes \mathbb{I})+\Pi_{A_{0}}^{3} \otimes (\sigma_x \otimes \sigma_x) \\
\end{split}
\label{v_4outcomes}
\end{equation}

\begin{figure}[t!]
\includegraphics[scale=0.6]{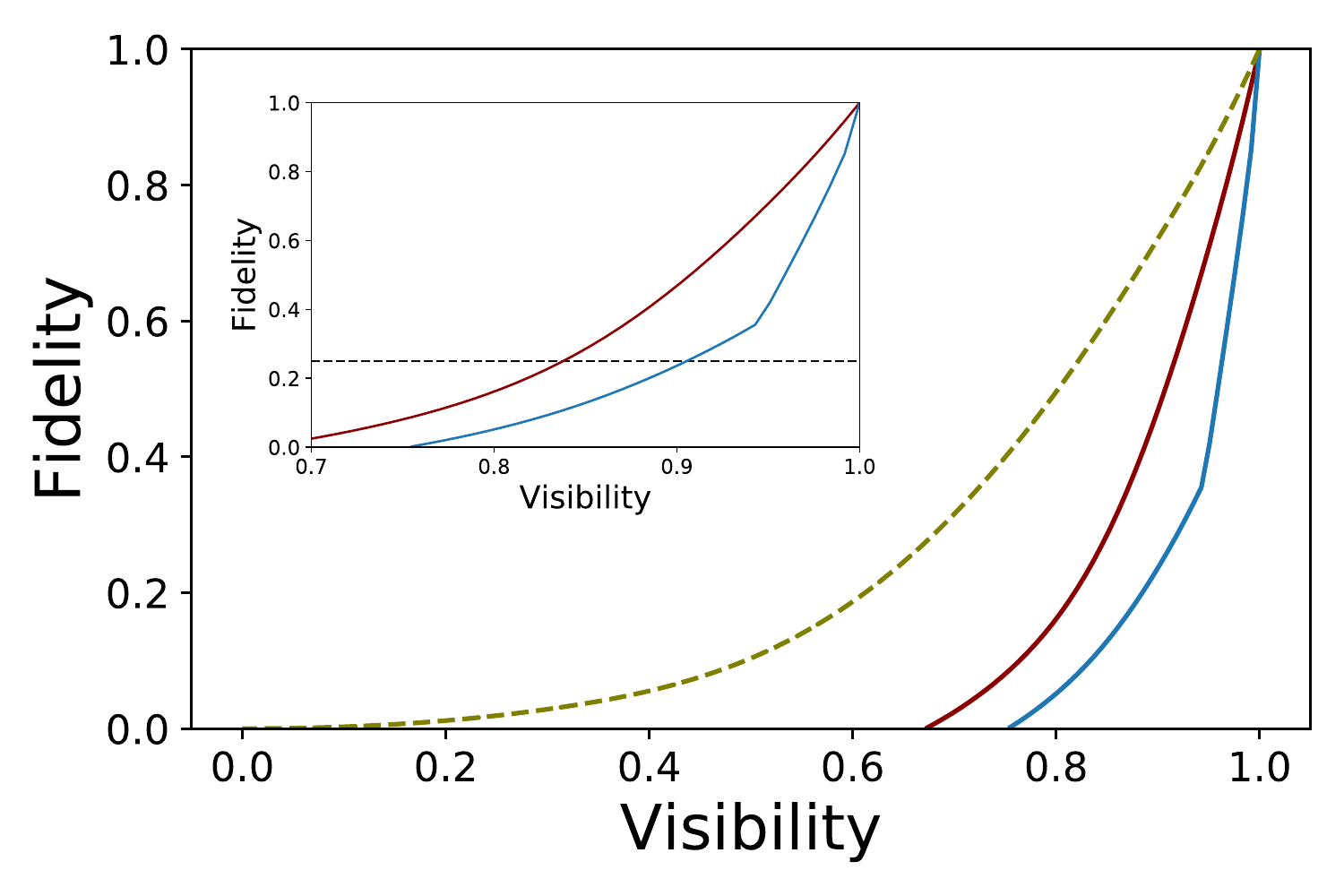}
\caption{\textbf{Certifiable fidelities in the bilocality and parallel self-testing scenarios.} 
The minimum fidelity certifiable with our protocol, respectively in the simple CHSH scenario (dashed curve) \cite{yang_2014}, in the bilocality scenario (red curve) and in the case of two  parties sharing two singlets, i.e. parallel self-testing scenario, (blue curve) \cite{supic_2019}. In the first case, the target state would be a 2-qubit maximally entangled state, $|\psi\rangle$, while in the other two cases, the target would be a product of two 2-qubit maximally entangled states, respectively $|\psi\rangle_{A_{1}B} \otimes |\psi\rangle_{A_{2}C}$ (bilocality scenario) and $|\psi \rangle_{A_{1}B_{1}} \otimes |\psi\rangle_{A_{2}B_{2}}$ (parallel self-testing scenario). The fidelities are plotted in terms of the visibilities of the generated states. In particular, we consider that, in the CHSH case, the generated state would have the following form: $\rho_{AB}= v |\psi \rangle \langle\psi|+(1-v) \frac{\mathbb{I}}{4}$. In the bilocality case, $\rho_{ABC}= \rho_{A_{1}B} \otimes \rho_{A_{2}C}$, with $\rho_{A_{1}B}= \rho_{A_{2}C}=$  $|\psi \rangle \langle\psi|+(1-v) \frac{\mathbb{I}}{4}$. In the parallel self-testing case, analogously, $\rho_{AB}= \rho_{A_{1}B_{1}} \otimes \rho_{A_{2}B_{2}}$, with $\rho_{A_{1}B_{1}}= \rho_{A_{2}B_{2}}=v|\psi \rangle \langle\psi|+(1-v) \frac{\mathbb{I}}{4}$. The inset highlights the bilocality and parallel self-testing cases and the black dashed line indicates the trivial fidelity.}
\label{fig:fidelities}
\end{figure}

After finding the complete expression of $\rho_{swap}$, the next step is to solve the semi-definite program defined in Eq.~\ref{npa_fidelity}, where probabilities $p(a_1,b_1|x,y)$ and $p(a_2,b_2|x,y)$ are given. For this scenario, it is sufficient to choose $\mathcal{Q}_n$ with $n=3$, i.e. NPA hierarchy \cite{NPA1, NPA2} at level 3, since the correlation term with highest degree appearing in the fidelity is $c_{A_0A_1A_0B_0B_1B_0} = Tr(\rho_{A_1B_1A_2B_2}A_0A_1A_0B_0B_1B_0)$.
The bound on the certifiable fidelity, considering a generated state $\rho_{A_{1,2}B_{1,2}}=\rho_{A_{1}B_{1}} \otimes \rho_{A_{2}B_{2}}$, where $\rho_{A_{1}B_{1}} = \rho_{A_{2}B_{2}}= v|\psi\rangle \langle \psi| +(1-v) \mathbb{I}/d$ is plotted in Fig.~\ref{fig:fidelities} as a function of the visibility $v$ (blue curve).

\subsection{The SWAP method for the tripartite scenario}
Let us now consider the case in which the states generated by sources $1$ and $2$ are sent to three parties (Alice, Bob and Charlie) that perform local measurements according to an input $(x,y,z) \in (0,1)$. In detail, Bob and Charlie perform 2-output local measurements on their subsystem, respectively generated by source $1$ and $2$, while Alice performs 4-output local measurements on the subsystem generated by both sources.
In such scenario, \textit{self-testing} could be achieved for a state product of two-qubit maximally entangled states, either performing two separate CHSH tests or a bilocality test \cite{Carvacho2017, andreoli_2017, saunders_2017}, when observing the maximum violation. However, if we violate the classical bound, but we do not observe the maximum violation extent or we want to self-test another state, these tests are not conclusive.
Therefore, we wish to bound the fidelity between the generated state and the target state, i.e.:
\begin{equation}
    F(\rho_{A_1BA_2C}, \ket{\psi_{target}})=\bra{\psi_{target}} \rho_{A_1BA_2C} \ket{\psi_{target}}
    \label{fidelity3parties}
\end{equation}
where the target is a state product of $\ket{\bar{\psi}}$, defined in Eq.~\ref{psi_bar}.
In order to test the state, the participants own an ancillary register each, with a dummy state of dimension $2$ for Alice, and $1$ for Bob and Charlie. 
As in the previous case, the parties perform the swap, and the resulting $\rho_{swap}$ density matrix will take the form:
\begin{equation}
    \rho_{swap}= Tr_{A_1BA_2C}[S(\rho_{A_1BA_2C}\otimes \ket{0000}\bra{0000})S^{\dag}] 
\end{equation}
The SWAP operator is defined as $S=S_{A_1A_1'} \otimes S_{BB'}  \otimes S_{A_2A_2'} \otimes S_{CC'} $, Alice's one having dimension $2$ and thus defined as in Eq.~\ref{u_4outcomes} and Eq.~\ref{v_4outcomes}, while Bob and Charlie's having dimension $1$ and thus defined as in Eq.~\ref{u} and Eq.~\ref{v}.

The $SDP$ finds a lower bound starting from the probabilities $p(a_1,b|x,y)$ and $p(a_2,c|x,z)$, but the computational power required is higher in this case, since  the correlation term with highest degree is $c_{A_0A_1A_0B_0B_1B_0C_0C_1C_0} = Tr(\rho_{A_1B_1A_2B_2}A_0A_1A_0B_0B_1B_0C_0C_1C_0)$, and level $5$ of the hierarchy would be necessary.
The bound on the certifiable fidelity, considering a generated state $\rho_{A_{1,2}BC}=\rho_{A_{1}B} \otimes \rho_{A_{2}C}$, where $\rho_{A_{1}B} = \rho_{A_{2}C}= v|\psi\rangle \langle \psi| +(1-v) \mathbb{I}/d$ is plotted in Fig.~\ref{fig:fidelities} as a function of the visibility $v$ (red curve).


\section{NPA hierarchy}
In this section we discuss the Navascués-Pironio-Ac{\'{\i}}n hierarchy, first introduced in \cite{NPA1}, which represents an efficient tool to solve optimization problems over the set $\mathcal{Q}$ of all quantum states and measurements of arbitrary dimension.


Let us consider the case of two parties, Alice and Bob, sharing a state $\rho$ and performing local measurements that, without loss of generality, we consider to be projectors, indicated with $E_{\alpha, \beta}$ respectively for Alice and Bob. Let us remind the properties of the projection operation: (\textit{i}) $\sum_{\mu}E_{\mu} = I$, (\textit{ii})  $E_{\mu}E_{\nu} = \delta_{\mu,\nu}$ for $E_{\mu}$ and $E_{\nu}$ belonging to the same measurement, and (\textit{iii}) $[E_{\alpha}, E_{\beta}] = 0$, namely projectors belonging respectively to Alice and Bob commute with each other. Now, let us assume that for a given probability distribution $P_{\alpha \beta}$ there exist a quantum state $\rho$ and a set $\{E_{\mu}  \}$ of projection operations such that 
\begin{equation}
    P_{\alpha \beta} = \Tr(\rho E_{\alpha} E_{\beta})
    \label{quantum_prob}
\end{equation}
The implications coming from this assumption give necessary conditions that must hold for the set $\{E_{\mu}  \}$. 

Indeed, let us consider the matrix $\Gamma$ defined as follows:
\begin{equation}
    \Gamma_{ij} = \sum_{ij} \Tr(S^{\dag}_i S_j \rho)
\end{equation}
where $\mathcal{S} = \{ S_1, S_2, ..., S_n \}$ is the set made by all the operators of the form $E_{\alpha}E_{\beta}E_{\alpha'}$ or $\sum_{\alpha}c_{\alpha}E_{\alpha}$. It can be shown \cite{NPA1} that for $\Gamma$, together with Hermitianity, the following linearity conditions hold:

\begin{equation}
    \sum_{ij} c_{ij}\Gamma_{ij} = 0, \quad \quad \mathrm{if~} \sum_{ij} c_{ij}S^{\dag}_iS_j = 0
    \label{lin1}
\end{equation}
\begin{equation}
    \begin{split}
        \sum_{ij} c_{ij}\Gamma_{ij} = \sum_{\alpha \beta} d_{\alpha \beta}P_{\alpha \beta}, \quad \quad
        \mathrm{if~} \sum_{ij} c_{ij}S^{\dag}_iS_j = \sum_{\alpha \beta} d_{\alpha \beta}E_{\alpha}E_{\beta}
    \end{split}
\label{lin2}
\end{equation}
and it is positive semidefinite:
\begin{equation}
    \Gamma \succeq 0
\label{SDP}
\end{equation}
It follows that, if for some set $\mathcal{S}$ it is not possible to find a $\Gamma$ matrix with the aforesaid properties, we must conclude that the probability distribution $P_{\alpha \beta}$ cannot be reproduced through local measurements on a quantum state. Additionally, it can be noted that if a set $\mathcal{S}$ can be written as linear combinations of operators in another set $\mathcal{S'}$, the conditions imposed by $\mathcal{S'}$ are at least as constraining as the ones imposed by $\mathcal{S}$, and that the set $\mathcal{S}_m$ of all possible products of $m$ projectors generates by linear combinations all the operators that are linear combinations of products of $m'$ projectors, with $m'\leq m$ \cite{NPA1}.

Hence, given a set of projectors $\{E_{\mu}\}$, the NPA method allows to build in a hierarchical way all the conditions, i.e. the sets $\mathcal{S}$, to be satisfied in order to certify the \textit{quantumness} of a given probability distribution. The hierarchy is based on constructing sets $\mathcal{S}_n$ made up of products of the given operators until degree $n$: for instance, we can first consider the set containing just the projectors of the two parties, Alice and Bob, that we indicate by $\mathcal{S}_1 = \{E_{\alpha}, E_{\beta} \}$. If conditions \eqref{lin1},  \eqref{lin2}, \eqref{SDP} hold, we can go on and verify the constraints imposed by the set $\mathcal{S}_2 = \{E_{\mu}E_{\nu} \}$, made up of all the products of the operators $\{E_{\alpha}, E_{\beta} \}$, then iterating the process until some condition fails, or until the set $\mathcal{S}_n$ is sufficiently large to represent all the conditions we need to check.

For a concrete example, let us consider Alice performing measurements $X = 1,2$ and Bob performing measurements $Y = 3,4$, where each measurement yields one of the two possible outcomes $\{a,b\} \in \{+1,-1\}$. Let us also define the correlation functions $C_{XY} = \sum_{ab} ab P(a,b|X,Y)$ and the marginal quantities $C_{X} = \sum_{a} a P(a,b|X,Y)$ and $C_{Y} = \sum_{b} b P(a,b|X,Y)$. The test of the NPA hierarchy corresponding to the first level, i.e. $n=1$, is built on the set $\mathcal{S}_1 = \{ X_1, X_2, Y_1, Y_2 \}$, and the $\Gamma$ matrix will have the following form:
\begin{equation}
    \Gamma=
    \begin{pmatrix}
        1 & C_1 & C_2 &  C_3   & C_4 \\
          &  1  &  u  & C_{13} & C_{14}\\
          &     &  1  & C_{23} & C_{24}\\
          &     &     &    1   & v\\
          &     &     &        & 1
    \end{pmatrix}
    \label{eq:gamma}
\end{equation}
where the lower symmetric part has been elided (remember that $\Gamma$ is hermitian), and the entries $u$ and $v$ correspond to the correlation terms involving non-commuting measurements (both belonging to Alice or Bob). These entries are thus indeterminate and can be adjusted by use of a semi-definite program in order to get an $SDP$ $\Gamma$ matrix, if the other correlations $\{ C_X, C_Y, C_{XY}\}$ are quantum \cite{NPA1}.


\subsection{Application to SDP characterization in parallel self-testing}
In the parallel self-testing scenario the function to be minimized, i.e. the fidelity defined in \eqref{fidelity2parties}, is a linear combination of correlation terms containing products of $2$ (one for Alice and one for Bob) up to $6$ (three each) operators. To minimize such function on a quantum set, we need to impose these terms to be the entries of one positive semi-definite $\Gamma$ matrix, and given that, as shown in the previous example, level $1$ of the hierarchy considers constraints on correlation terms up to degree $2$, in our case it is necessary to build the hierarchy up to level $3$, at least. To do so, we exploit the Python library first introduced in \cite{wittek_2015}, which allows to set both the objective function to be optimized and the level of the relaxation, and then to solve the SDP problem with MOSEK optimization software \cite{mosek}.

Now, considering that Alice and Bob own $8$ projector operators each, indicated with $\Pi^a_{A/B_x}$, where $a=\{0,1,2,3\}$ are the possible eigenvalues, and $x \in (0,1)$, and given also property (\textit{i}) of the projection operation, yielding $\Pi^3_{A/B_x} = \mathbb{I} - (\Pi^0_{A/B_x} + \Pi^1_{A/B_x} + \Pi^2_{A/B_x})$, at level $1$ the $\mathcal{S}_1$ set contains the following $12$ elements $\mathcal{S}_1 =\{ \Pi^0_{A_0}, \Pi^1_{A_0}, \Pi^2_{A_0}, \Pi^0_{A_1}, \Pi^1_{A_1}, \Pi^2_{A_1}, \\ \Pi^0_{B_0}, \Pi^1_{B_0}, \Pi^2_{B_0}, \Pi^0_{B_1}, \Pi^1_{B_1}, \Pi^2_{B_1}  \}$. Given also that the number of the $\Gamma$ entries grows with the level of the relaxation and with the number of operators involved, and decreases with the number of allowed substitutions, namely if entry $\Gamma_{ij} = \Pi^0_{A_0} \times \Pi^0_{A_0} \times \Pi^0_{B_1}$ and entry $\Gamma_{i'j'} = \Pi^0_{A_0} \times \Pi^0_{B_1}$, than the two are the same variable, due to property (\textit{ii}) of the projection operation, the total number of variables concerned in this optimization problem is found to be $22602$.

However, it is possible to further reduce this quantity by noticing that our objective function is symmetric under exchange of parties $A \leftrightarrow B  $, and that this transformation keeps also the matrix $\Gamma$ unchanged. In particular, we added to the aforementioned substitutions those of the kind:

\begin{equation}
    \begin{split}
    \Pi^a_{A_0} &: \Pi^a_{B_0} \\
    \Pi^a_{A_1} &: \Pi^a_{B_1} \\
    \Pi^a_{A_0} \times \Pi^{a'}_{A_1} &: \Pi^a_{B_0} \times \Pi^{a'}_{B_1} \\
    \Pi^a_{A_0} \times \Pi^{a'}_{B_1} &: \Pi^{a'}_{A_1} \times \Pi^{a}_{B_0} \\
    \Pi^a_{A_0} \times \Pi^{a'}_{A_1} \times \Pi^{a''}_{B_0} &: \Pi^{a''}_{A_0} \times \Pi^a_{B_0} \times \Pi^{a'}_{B_1} \\
    ...
    \end{split}
\end{equation}

where we took care to respect the commutation rules of Alice and Bob's operators when defining LHS and RHS of each substitutions.

\subsection{Application to SDP characterization in the tripartite scenario}
Let us now consider the scenario in which Alice shares one singlet with Bob and another with Charlie. Here, the fidelity to be bounded is the one in \eqref{fidelity3parties}, which contains correlation terms of at least $3$ (one for each party), and at most $9$ (three for each party) operators. In this case, the level $1$ of the hierarchy would be built over the following set $\mathcal{S}_{bilo} =\{ \Pi^0_{A_0}, \Pi^1_{A_0}, \Pi^2_{A_0}, \Pi^0_{A_1}, \Pi^1_{A_1}, \Pi^2_{A_1},
\Pi^0_{B_0}, \Pi^0_{B_1}, \Pi^0_{C_0}, 
\Pi^0_{C_1} \}$,
containing $10$ projection operators, then at the minimum required level, that is $5$, the problem would be too computationally requiring for a normal computer. To simplify the problem, it is possible to stop at level $3$ of the hierarchy, though considering all the monomials with degree higher than $6$ appearing in the objective function as \textit{extramonomials}. In this way, not all the possible product combinations of the operators are taken into account as variables, for correlation terms of degree $ > 6$, but only those given as \textit{extramonomials}. The number of variables concerning the optimization is, considering also the standard substitutions due to the projection properties, is $10115$. Besides this, it is possible to further reduce the number of the $\Gamma$ entries by using, again, the symmetries of the objective function. This time, the symmetry holds for the exchange of the peripheral parties, i.e. Bob and Charlie, $B \leftrightarrow C$ , and swapping the second and third output of the central party, i.e. $01 \leftrightarrow 10$. By considering this symmetry when setting the list of the substitutions, the number of SDP variables decreases to $7670$, while the time for the optimization decreases from $136.28 s$ to $78.16 s$.

\section{Frequency regularization and definition of the dual equality constraint}
\label{sec:regularization}
Navascués, Pironio and Ac\'{i}n hierarchy is just one of the several examples of theoretical tools taking into account the full quantum distribution for device-independent characterization, which all share the common assumption of non-signaling. However, due to finite statistics, raw distributions obtained by the experimental frequencies generally do not satisfy this condition, to the extent that it is almost always impossible to use raw data within these methods. In \cite{lin2018device}, a general tool called the \textit{device-independent least-square method} is introduced, which can be used to estimate the non-signaling probability distribution, belonging to a superset of the quantum correlation set, $\mathcal{Q}_l$, which is the closest to the experimental frequencies, in terms of a norm-2 distance ($|| \vec{f} - \vec{P} ||_2$). In particular,  we indicate with $\vec{f}$ the experimental frequencies and $P$ is the probability distribution satisfying the no-signalling condition. Let us briefly remind the non-signaling condition: a distribution of probability $P(a,b|x,y)$ is non-signaling if the following two relations are simultaneously satisfied:
\begin{equation}
\begin{split}
    P(a|x,y) &\equiv \sum_b P(a,b|x,y) = P(a|x,y'), \quad \forall a,x,y,y' \\
    P(b|x,y) &\equiv \sum_a P(a,b|x,y) = P(b|x',y), \quad \forall b,y,x,x'
\end{split}
\end{equation}

In detail, this technique is shown in \cite{lin2018device} to be equivalent to performing a projection $P_{\Pi}(\vec{f})$ of $\vec{f}$ onto an affine subspace $\mathcal{N}$ of $\mathbb{R}^{16}$ ($16$ is the dimension of the frequency vector in a simple CHSH 2-parties scenario), which contains only $\vec{P}$s satisfying the non-signaling condition, followed by the minimization of the 2-norm distance between $P_{\Pi}$ and $\mathcal{Q}$.

Formally, the method amounts to finding the unique minimizer of the least-square problem:
 \begin{equation}
     \vec{P}_{LS}(\vec{f}) = \argmin_{\vec{P} \in \mathcal{Q}} || \vec{f} - \vec{P} ||_2
     \label{eq:least_square}
 \end{equation}
Where, when the quantum set $\mathcal{Q}$ is approximated by a superset relaxation $\mathcal{Q}_l$ that admits a semi-definite programming characterization, through the NPA hierarchy.
Problem \eqref{eq:least_square} can be cast as an SDP, although the norm-2 distance is not a linear function, in the following formulation, using the characterization of positive semidefinite matrices via their Shur's complement:
\begin{equation}
    \begin{split}
        & \argmin_{\vec{P} \in \mathcal{Q}_l} \quad \quad s \\
        & \text{s.t.}
        \begin{pmatrix}
        s\mathbb{I} & \vec{f} - \vec{P} \\
        \vec{f}^T - \vec{P}^T & s
        \end{pmatrix}
        \succeq 0 \\
        & \text{and } \Gamma \succeq 0
    \end{split}
\end{equation}
where $\mathbb{I}$ is the identity matrix having the same dimension of the column vector $\vec{f}$, and $\Gamma$ is defined as in Eq.\eqref{eq:gamma}, depending on the chosen NPA hierarchy level. 
Indeed, by Theorem 7.7.7 of \cite{horn2012matrix}, 
\begin{equation}
    s- (\vec{f}^T-\vec{P}^T) \frac{\mathbb{I}}{s} (\vec{f}-\vec{P}) \succeq 0
    \label{eq1}
\end{equation}
given that $(\vec{f}^T-\vec{P}^T)(\vec{f}-\vec{P})=||\vec{f} - \vec{P} ||^2_2$, Eq.~\eqref{eq1} is equivalent to 
\begin{equation}
   s^2 \succeq  ||\vec{f} - \vec{P} ||_2^2
    \label{eq2}
\end{equation}
and hence $s \succeq  ||\vec{f} - \vec{P} ||_2$.

This problem only involves an objective function and matrix constraints that are linear in the SDP variables $s$, $\vec{P}$ and in the variables of the $\Gamma$ matrix, and is shown \cite{lin2018device} to be unique and totally equivalent to the one defined is Eq.~\eqref{eq:least_square}.

At this point, the obtained probability distribution $\vec{P}$ is injected in the SDP program as linear constraints on the moments of the $\Gamma$ matrix, as described in Eq.~\eqref{npa_fidelity} of the Supplementary information or Eq.~(10) of main text, where the $p$ are the regularised probabilities.

Now, consider that given $C \in \mathcal{M}^n$, $A_i \in \mathcal{M}^n$, $i=1,2,...,m$, and $b \in \mathcal{R}^m$, the semidefinite programming problem is to find a matrix $X \in \mathcal{M}^n$ for the following optimization problem:
\begin{equation}
    \begin{split}
    \text{inf} \quad \quad \quad &C \cdot X \\
    \text{subject to    } &A_i \cdot X = b_i ~|~ i=1,...,m \; | ~X \succeq 0
    \end{split}
\end{equation}
which individuates the \textit{primal} solution to the problem, i.e. searches for the minimum solution from above. The corresponding dual problem can be written as:
\begin{equation}
\begin{split}
    \text{sup} \quad \quad \quad &b^Ty \\
    \text{subject to      } &\sum_i y_iA_i + S = C \; |~S \succeq 0
\end{split}
\end{equation}
Here, $y \in \mathcal{R}^m$ and $S \in \mathcal{M}^n$ constitute the solution of the dual problem, namely the best approximation of the result from below, which coincides with the primal, due to strong duality. Note that the objective function of the dual problem, $b^Ty$, is, in our case, a linear combination of the regularized frequencies $\sum_i^m c_i P_i$. 
At this point, we took the dual solution of the SDP, namely, the vector of coefficients $y$, and evaluated the linear combination $D_{exp} = b_{exp}^T y$ by putting as $b_{exp}$ the observed frequencies. In the end, this new linear combination of the matrix moment is given as equality constraint to a new SDP which finally gives our device-independent estimation of the minimal certifiable fidelity, as follows:
\begin{equation}
    \begin{split}
    f=& \min~\bra{\psi_{target}}\rho_{swap}\ket{\psi_{target}}\\
   \textrm{s.t.} ~ & c \in \mathcal{Q}_l,\\
    & c^T y = b_{exp}^T y
    \end{split}
\label{final-sdp}
\end{equation}
This method allows to extract an experimental lower bound on the fidelity, directly from the experimental data and not from the regularised one, avoiding a possible overestimation of it.

\section{Bounding the confidence level on the fidelity by use of Hoeffding's inequality}
Hoeffding's inequality \cite{hoeffding_1963} represents a very useful tool when there is need to estimate uncertainties on experimental probabilities, taking into account finite statistics. It asserts that given $n$ independent random variable $X_1,...,X_n$, defined in $[0,1]$, with the following mean, expected value and variance:
\begin{equation*}
\begin{split}
    S &= (X_1 + X_2 + ... + X_n) \\
    \overline{X} &= S/n \\
    \mu &= E\overline{X} = ES/n \\
    \sigma^2 &= \Var(S)/n
\end{split}
\end{equation*}
Then, the following inequality holds, for $0 \leq t \leq 1-\mu$:
\begin{equation}
    \epsilon = Pr(\overline{X}-\mu \geq t) \leq e^{-2nt^2}
    \label{eq:hoeff1}
\end{equation}
where $t$ depends on the arbitrary confidence interval $\epsilon$ assigned to the variable $y = \overline{X}-\mu$ according to: $t(\epsilon) = \sqrt{-\ln(\epsilon)/(2n)}$.

From this bound, it is possible to extract information about the effects of finite statistics on the estimation of the dual constraint, described in \ref{sec:regularization}. Let us consider, in fact, the worst-case scenario of Hoeffding's inequality \eqref{eq:hoeff1}, i.e. $\epsilon = e^{-2nt^2}$. The RHS of this equality corresponds to one minus the cumulative distribution of the variable $y$, namely:
\begin{equation}
    P\{y \geq t\} = \int_{t}^{\infty} P(y)dy= e^{-2nt^2}, \quad \quad \quad t>0 
    \label{eq:cumulative}
\end{equation}
\begin{figure}[ht!]
    \centering
    \includegraphics[width=0.5\linewidth]{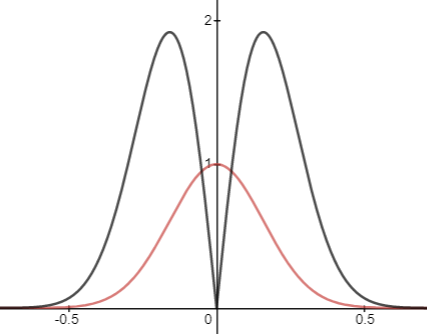}
    \caption{\textbf{Hoeffding's distribution function}. The black curve represents the normalised probability distribution of the variable $y = \overline{X}-\mu>0$, $P(y) = 2n \left|y\right| e^{-2ny^2}$, while the red curve represents 1 minus the cumulative probability of such a distribution, i.e. the probability that $y$ lies outside the range $[0,t]$, given by Hoeffding's inequality \cite{hoeffding_1963}. Since, for the case $y < 0$, Hoeffding's inequality holds simmetrically and, hence, $P(y)=P(-y)$, so, the probability that y lies outside the range $[-t,t]$ amounts to $P(\left| y \right| \geq t, t>0) = \int_{-\infty}^{-t} -2n y e^{-2ny^2} dy + \int_t^{\infty} 2n y e^{-2ny^2} dy = e^{-2nt^2}$. The variance of the black function is $Var(y) =  1/(2n)$. We used the variance of this distribution to compute the statistical uncertainty on the constraint $c^T(y)$ of problem \eqref{final-sdp}, by asserting that it lies inside the interval $d_{exp}-\tau(\epsilon) \leq c^Ty \leq d_{exp}+\tau(\epsilon) $ with a probability that can be recovered from a standardized normal table, considering that such interval amounts to $\sqrt{-\ln(\epsilon)}$ standard deviations.}
    \label{fig:cumulative}
\end{figure}
For the Fundamental Theorem of Calculus, the probability distribution function $P(y)$ for $y>0$ can be obtained by differentiating one minus \eqref{eq:cumulative}, thus yielding to:
\begin{equation}
    P(y) = \frac{d (1-e^{-2ny^2})}{dy} = 4nye^{-2ny^2}
    \label{eq:probability}
\end{equation}
which is a well-defined, positive and normalized distribution of probability, when considering $y>0$. For the case $y<0$, Hoeffding's inequality still holds symmetrically \cite{hoeffding_1963}, leading to the following an upper bound, for $t>0$:
\begin{equation}
    Pr(-\overline{X} + \mu \geq t) = e^{-2nt^2}
\end{equation}
This means that we can consider the two-sided variant of Hoeffding's bound, relative to the absolute value of the variable $y$, i.e.:
\begin{equation}
    Pr(\left| \overline{X}  - \mu \right| \geq t) = 2e^{-2nt^2}
    \label{eq:abs_cumulative}
\end{equation}
by defining normalized probability distribution over all real values of $y$, which is $P(y) = 2n\left| y \right |e^{-2ny^2}$ and whose variance is defined as $Var(y) = \frac{1}{2n}$.
For the central limit theorem, a linear combination $\sum_{k=1}^N a_k y_k$ of $N$ such independent, finite variance variables has a Gaussian distribution characterised by the following variance:
$$\sigma_{gauss}^2 = \sum_{k=1}^N a_k^2 (\Var[y_k]) = \sum_{k=1}^N a_k^2/(2n)$$
Since the experimental value of the dual $d_{exp}$ is, actually, a linear combination of the experimental frequencies, its probability distribution can be approximated by a Gaussian, and we can estimate that the experimental dual lies within the following interval:
\begin{equation}
    d_{exp}(f) - \tau(\epsilon) \leq d(p) \leq d_{exp}(f) + \tau(\epsilon)
\end{equation}
where
\begin{equation}
     \tau^2(\epsilon) = \sum_{k=1}^N a_k^2t^2_k(\epsilon) = -\ln(\epsilon) \sigma^2
\end{equation}
having defined $\sigma^2 = \sum_{k=1}^N a_k^2 \Var(y_k)$, with a confidence level that can be computed from a standardized normal table considering that such interval amounts to $\sqrt{-\ln(\epsilon)}$ standard deviations.